\documentclass[tightenlines,nofootinbib,preprint,aps,amsmath,superscriptaddress]{revtex4}
\usepackage{graphicx}
\usepackage{bm}
\usepackage{epsfig}
\usepackage{amsmath,latexsym}

\def\Dslash{D\!\!\!\!\slash}
\def\pslash{p\!\!\!\slash}

\def\OMIT#1{}

\newcommand{\nn}{\nonumber}

\newcommand{\bea}{\begin{eqnarray}}
\newcommand{\eea}{\end{eqnarray}}

\begin{document}

\preprint{\tighten\vbox{\hbox{UCB-PTH-03/06}\hbox{CMU-HEP-03-02}\hbox{UCSD-PTH-03-02}\hbox{hep-ph/0303158}}}

\title{\phantom{x}\vspace{2cm}
Systematics of Coupling Flows in AdS Backgrounds}

\author{Walter D. Goldberger}\email{walter@thsrv.lbl.gov}
\affiliation{Department of Physics, University of California,
Berkeley, CA 94720\vspace{6pt}}
\affiliation{Theoretical Physics Group, Lawrence Berkeley National
Laboratory, Berkeley, CA 94720\vspace{6pt}}
\author{Ira Z. Rothstein}\email{ira@cmuhep2.phys.cmu.edu}
\affiliation{Department of Physics, Carnegie Mellon University,
Pittsburgh, PA 15213\vspace{6pt}}
\affiliation{Department of Physics, University of California at San Diego, La Jolla, CA 92037}

\begin{abstract}
\vspace{0.5cm} \setlength\baselineskip{18pt} 

We give an effective field theory derivation, based on the running of Planck brane gauge correlators, of the large logarithms that arise in the predictions for low energy gauge couplings in compactified $\mbox{AdS}_5$ backgrounds, including the one-loop effects of bulk scalars, fermions, and gauge bosons.  In contrast to the case of charged scalars coupled to Abelian gauge fields that has been considered previously in the literature, the one-loop corrections are not dominated by a single 4D Kaluza-Klein mode.  Nevertheless, in the case of gauge field loops, the amplitudes can be reorganized into a leading logarithmic contribution that is identical to the running in 4D non-Abelian gauge theory, and a term which is not logarithmically enhanced and is analogous to a two-loop effect in 4D.  In a warped GUT model broken by the Higgs mechanism in the bulk, we show that the matching scale that appears in the large logarithms induced by the non-Abelian gauge fields is $m_{XY}^2/k$ where $m_{XY}$ is the bulk mass of the $XY$ bosons and $k$ is the AdS curvature.  This is in contrast to the UV scale in the logarithmic contributions of scalars, which is simply the bulk mass $m$.   Our results are summarized in a set of simple rules that can be applied to compute the leading logarithmic predictions for coupling constant relations within a given warped GUT model. We present results for both bulk Higgs and boundary breaking of the GUT gauge group.

\end{abstract}
\maketitle


\newpage


\section{Introduction}

Recent progress in understanding renormalization in compactified anti-deSitter space (AdS) has led to the intriguing possibility of constructing models which preserve grand unification at an exponentially large scale, and at the same time realize the physics of extra dimensions at the TeV scale. These models are variations of the Randall Sundrum (RS) scenario~\cite{RS1} in which two branes  act as boundaries to a bulk region of 5D AdS space. In conformal coordinates the spacetime geometry takes the form
\begin{equation}
ds^2={1\over (kz)^2} \left(\eta_{\mu\nu} dx^\mu dx^\nu-dz^2\right),
\end{equation}
where $z$ parametrizes location in the bulk spacetime, bounded by a UV (Planck) brane at $z=1/k$ and an IR (TeV) brane at $z=1/T$. $x^\mu$ labels 4D Poincare coordinates parallel to the branes.  The curvature scale $k$ is taken to be slightly below the fundamental scale $M_5\sim M_{Pl}$.  Despite the fact that the proper distance between the two branes is order $1/k$, the masses of the low-lying Kaluza-Klein (KK) states scale with $T$, which to address the hierarchy problem is taken to be of order the TeV scale (a simple mechanism for the generation of this exponentially low energy scale can be found in~\cite{gw}).

Given that the structure of KK states propagating in this background is similar to the 4D spectrum of a bulk field propagating in 5D flat space compactified on an interval of radius $R\sim(\mbox{TeV})^{-1}$, one would imagine that a low energy observable such as the weak mixing angle receives large quantum corrections, of order $M_{GUT}/T$, from GUT scale physics.  In this case, the standard tools of perturbative field theory can be effectively applied only for relative low values of $M_{GUT}$ (a loop factor above $T$, say).  Even then, the linear dependence of loop corrections on the values of heavy particle masses introduces a loss of predictive power.

However, loop corrections to low energy observables in compactified
AdS actually scale as a log of $q/M_{GUT}$, with $q$ a
low energy scale typically of order the weak scale. This possibility
was first raised in~\cite{pomarol} for the case of Abelian gauge
fields coupled to scalar matter fields.  Perhaps the most
straightforward way to obtain this result is to calculate the low
energy vacuum polarization graphs, summing over the tower of KK modes in the
loops. While technically correct, the calculation is cumbersome and
not terribly illuminating, as it gives no clue as to how this miracle
occurs.  Instead, in~\cite{GnR1} we showed how the one-loop logarithms in the
low energy coupling predictions emerge naturally from the study of
correlators which are restricted to the Planck brane and have external
momenta greater than $T$.  Due to the relation between energy scale
$p_4$ and bulk position $z$ in background AdS geometries, $p_4\sim
M_{Pl}/(kz),$ such correlators are guaranteed to remain perturbative
up to energies much higher than the KK mass scale.  Yet at low
energies they must match directly onto the measured zero mode
observables.  Thus by following their flow across a wide range of
energies, one can use these Planck correlators to correctly account
for the dependence of low energy quantities on UV scales.  The
remarkable fact that such flows turn out to be logarithmic, as in a
purely 4D theory, is perhaps not surprising given that the AdS/CFT
correspondence~\cite{maldacena,AdSCFT}, as it applies to RS
scenarios~\cite{Gubser,APR,RZ,PV}, implies that there is a well defined
prescription for relating Planck correlators in the AdS background to
quantities in a 4D conformal field theory (CFT).  The non-decoupling
approach to computing logarithmic corrections can be found
in~\cite{ads,cct,GnR2,choidec,ads1}.  For other approaches to understanding
the logarithms that arise in the low energy predictions,
see~\cite{rschw,choi,decon,ad}.

Not only is the momentum dependence of the Planck correlators logarithmic, but
the precise value of the numerical coefficient of the non-universal
one-loop logarithms can be unambiguously calculated~\footnote{Note
that in AdS, the two-point Planck correlator of a gauge field exhibits
logarithmic running with energy even at tree-level.  In a GUT context,
such logarithms are completely universal and therefore do not enter
into the prediction for the weak mixing angle.}.  In~\cite{GnR1}, we
used AdS/CFT arguments to show that the coefficient of the one-loop
logarithm induced by a bulk charged scalar field coupled to a $U(1)$
gauge field is identical to what one finds in a 4D theory.  There we
also gave a simple argument for understanding this logarithmic flow
purely in AdS language, in terms of the structure of the KK
wavefunctions of bulk fields.  We elaborated on this line of reasoning in~\cite{GnR2}.  Essentially, all KK modes of a scalar field with
masses below $k$ are localized toward $z=1/T$ (that is, their
wavefunction is exponentially smaller at $z=1/k$) except for a single
zero mode whose overlap with the Planck brane is of order unity.
It follows that the effect of a bulk scalar field on the Planck brane
vacuum polarization is summarized simply by the loop contribution of
this zero mode, i.e., the computation reduces to a purely 4D one-loop
integral. 

These arguments naively break down when bulk gauge bosons or fermions
appear in the loops.  In such situations, it is no longer always true that
the KK resonances have exponentially small amplitudes on the Planck
brane relative to the zero mode.  We will show that despite the fact
that a large number of modes contributes to the two-point function, the
relative running of couplings in non-Abelian bulk gauge theories, as
measured by Planck observers, is identical to what one obtains by
solving renormalization group (RG) equations in ordinary 4D gauge theories to leading log
accuracy.  The relevant analysis can be found in
Section~\ref{sec:gauge}.  To make the discussion self contained, we
first review in Section~\ref{sec:scalar} the analysis of scalar loop
contributions to the running of the gauge couplings.  Bulk fermions
possess zero modes for any value of the bulk mass, but can lead
to incalculable corrections if the mass is larger than the curvature scale $k$.  These issues
as well as other subtleties related to bulk fermions are discussed in
Section~\ref{sec:fermions}. We conclude the paper with a section which
is meant as a tool kit for model building. Those who are not
interested in the technical issues will find in the final section a
summary of the relevant aspects of our analysis.

\section{Large logarithms from Planck correlators}
\label{sec:planck}

We take the point of view that any large logarithm that appears in a
low energy quantity must arise from the RG flow of an observable that
is well defined in the UV.  Because of the correlation between bulk
position and energy scales implied by the AdS metric, the only observables that are truly accessible to effective field
theory methods up to energies near the Planck scale are bulk field
Green's functions whose external points are localized near the region
$z=1/k$ of the compactified AdS background.  Thus in \cite{GnR1} we
chose to work directly with correlators whose endpoints lie strictly
on the Planck brane, at $z=1/k$.  In particular, we define the running
coupling for a gauge theory in AdS in terms of the two-point function
of the bulk gauge field
\begin{equation}
\Delta_{\mu\nu}(q^2)\equiv\int d^4 x e^{iq\cdot x} \langle A_\mu(x,1/k) A_\nu(0,1/k)\rangle.
\end{equation}
At least in a theory with $U(1)$ gauge symmetry (we discuss later the extension of this program to non-Abelian groups), this gives rise to an effective coupling
\begin{equation}
\Delta_{\mu\nu}(q^2)= {g^2(q)\over q^2}\eta_{\mu\nu} +\mbox{gauge},
\end{equation}
where we denote $q=\sqrt{q^2}$.

Before considering one-loop corrections, let us review the computation of $g^2(q)$ at tree level.  For simplicity, we will work in Euclidean signature throughout.  Using the gauge boson propagator (in Landau gauge) from Appendix~\ref{app:props}, one finds
\begin{equation}
\label{eq:tree}
{g^2(q)\over q^2} = g_5^2 D^{(v)}_q(1/k,1/k) = {g_5^2\over q} {K_1(q/k)\over K_0(q/k)}. 
\end{equation}
Note that the expression for the propagator we have used to define the running coupling $g(q)$ is insensitive to the effects of the brane at $z=1/T$.  This is adequate as long as we are probing the gauge potential at momenta $q\gg T$, for which the effects of the TeV brane are irrelevant.   We will also restrict ourselves to momenta less than the scale $k$.  For momenta larger than $k$, the Green's functions only probe field correlations over distances that are so short that one may as well neglect the effects of curvature and compute directly in a flat background space.  So for $T\ll q\ll k$, Eq.~(\ref{eq:tree}) gives rise to 
\begin{equation}
\label{eq:classrun}
g^2(q) \simeq {g_5^2 k\over\ln(2k/q)}.
\end{equation}
This is  the familiar classical running of the gauge couplings as seen by Planck brane observers.  A discussion of the origin of this effect from the point of view of the AdS/CFT correspondence can be found in~\cite{APR}.  

Already at this level, this result gives a simple illustration of the general philosophy that all large logarithms in low energy quantities arise as a result of running Planck brane correlators down to low energy.  Consider the gauge coupling measured by a low energy observer.  Such an observer only probes the interactions of the lightest charged KK states with the KK zero mode of the bulk gauge field.  In terms of the 5D coupling $g_5$, the leading order result for the coupling of these light states to the gauge field zero mode follows straightforwardly from dimensional reduction of the 5D action
\begin{equation}
{1\over g_4^2} = {R\over g_5^2},
\end{equation}
where $R$ is the separation between the UV and the IR brane.  In terms of our parametrization of the 5D metric, this is given by 
\begin{equation}
R={1\over k}\ln\left({k\over T}\right),
\end{equation}
which involves a large logarithm of the ratio of two energy scales.  On the other hand, consider the gauge coupling measured at low energies by observers localized to the $z=1/k$ boundary.     To predict the result of such a measurement, one should use the running Planck brane coupling evaluated at momenta $q\sim T$.  But  according to Eq.~(\ref{eq:classrun}), this quantity is in agreement (up to small uncertainties  in the precise value of the matching scale)  with the zero mode result.  It must be, since momenta which are less than the KK mass gap $T$ cannot distinguish between Planck observers and observers composed of KK zero modes of bulk fields.  It follows that the large logarithms that arise in low energy quantities can be obtained from the flow toward the IR of running couplings defined in terms of Planck correlators.  We now turn to the computation of quantum corrections to Planck brane correlators due to bulk scalar fields, non-Abelian gauge boson loops, and fermion fields.

\subsection{Scalars}
\label{sec:scalar}
Let us first review the computation of the one-loop vacuum polarization effects  of a charged scalar coupled to a $U(1)$ gauge field along the lines of~\cite{GnR1,GnR2}.  Following the notation for the propagators discussed in the appendix, we find at one-loop
\begin{equation}
\Delta_{\mu\nu}(q^2) = g_5^2 P_{\mu\nu}(q) D^{(v)}_q(1/k,1/k) + g_5^4 \left(L^{(1)}_{\mu\nu}(q^2) + L^{(2)}_{\mu\nu}(q^2)\right),
\end{equation} 
where $P_{\mu\nu}(q)= \eta_{\mu\nu} - q_\mu q_\nu/q^2$,
\begin{eqnarray}
\label{eq:L1}
\nn
L^{(1)}_{\mu\nu}(q^2) &=&\int {d^D p\over (2\pi)^D}(2p+q)_\mu (2p+q)_\nu\\
& & {}\times \int{dz\over (kz)^{D-1}}{dz'\over (kz')^{D-1}} D^{(v)}_q(1/k,z) D^{(s)}_p(z,z')D^{(s)}_{p+q}(z,z')D^{(v)}_q(z',1/k), 
\end{eqnarray}
is the usual vacuum polarization graph, and
\begin{equation}
\label{eq:L2}
L^{(2)}_{\mu\nu}(q^2)=-2\eta_{\mu\nu}\int{d^D p\over (2\pi)^D} {dz\over (kz)^{D-1}} D^{(v)}_q(1/k,z)  D^{(s)}_p(z,z)D^{(v)}_q(z,1/k).
\end{equation}
is the seagull term.  The scalar field propagator is denoted by $D^{(s)}_p(z,z')$.  In both these equations, we have for simplicity dropped purely longitudinal components of the gauge boson propagators, since these are not relevant to our definition of the running coupling.  Also note that  we will regulate possible UV divergences that may arise by working in dimensional regularization and subtracting poles at $D=4$.

One can quantitatively determine the leading momentum dependence of these integrals in the region $T\ll q\ll k$  by the following arguments.  First note that 
\begin{equation}
D^{(v)}_q(1/k,z) = {1\over q}(kz)^{D/2-1} {K_{D/2-1}(qz)\over K_{D/2-2}(q/k)}
\end{equation}
is exponentially suppressed away from from the region $qz\leq 1$ due to the asymptotic behavior of the modified Bessel function, $K_\nu(z)\sim\sqrt{\pi/2z}\exp(-z)$ for $z\gg 1$.  Therefore the integrals over $z,z'$ in Eq.~(\ref{eq:L1}) and Eq.~(\ref{eq:L2}) have support in the region $1/k<z<1/q$.  However, since $K_\nu(z)\sim z^{-\nu}$ for $z\ll 1$ the bulk of the contribution to the integral over this region arises from values of $z$ near $1/k$,  where $qz$ is smallest\footnote{ In the region where $z\approx 1/q$, there is no suppression from the external lines. However, in this region the integrand has no poles that could give rise to logarithmic behavior and is well behaved at $q=0$. Thus this contribution is power suppressed.}.  We conclude from this that only the behavior of the integrand near $z=1/k$ is important for the determination of the one-loop running.  

The loop amplitudes can be further simplified by noting that for
external momenta $q\ll k$, the leading non-analytic behavior of the
one-loop amplitude can be obtained by replacing all internal loop
propagators $D_l(z,z')$ with their expansions in the limit $lz,lz'\ll
1$ (here we denote by $l$ any combination of the external momentum $q$
and the loop momentum integration variable $p$).  Because the external
gauge boson lines force the range of $z,z'$ to effectively lie near
$1/k$, this approximation clearly breaks down for loop momenta of
order $k$.  However, for $q/k\ll 1$, loop momenta above the scale $k$
cannot give rise to the non-analytic effects that we are trying to
determine.  Thus our approximation is an effective choice of scheme,
since the contributions that have been dropped are analytic in the
external momentum.  This choice of scheme, if used consistently
throughout, cannot affect the low energy predictions for physical
observables.

To make this more explicit, we use the $pz,pz'\ll 1$ asymptotic expansion 
\begin{equation}
\label{eq:asymp}
D^{(s)}_p(z,z')\simeq (kz)^{D/2-\nu} (kz')^{D/2-\nu}{2k(\nu-1)\over p^2 +\left(1-{2\over D}\right) m^2},
\end{equation} 
of the propagator of a scalar field with mass $m$ (the full scalar propagator can be found in the appendix).  In this equation, $\nu=\sqrt{(D/2)^2 + m^2/k^2}$, and we have taken $m/k\ll 1$.  Note that the pole structure in this propagator suggests that our loop amplitude receives a dominant contribution from a single 4D mode.  For $m=0$, both the location of the pole in Eq.~(\ref{eq:asymp}) as well as the residue can be obtained from the KK mode expansion of the propagator
\begin{equation}
\label{eq:KKsum}
D^{(s)}_p(z,z') = \sum_n {\psi_n(z)\psi_n(z')\over p^2+m_n^2},
\end{equation}
(with $\psi_n(z)$, $m_n$ the $n$-th KK wavefunction and mass respectively) by keeping only the KK zero mode of the bulk scalar, with $m_0=0$ and $\psi_0(z)=\sqrt{k(D-2)}$.  It is perhaps surprising, given that we are expanding the propagator in a region of momentum space $p^2\gg T^2$, that effectively only the zero mode contributes to the sum in Eq.~(\ref{eq:KKsum}).  The dominance of the zero mode is in fact a consequence of the background AdS curvature, which causes an exponential suppression of the excited KK state wavefunctions relative to the $n=0$ mode near $z=1/k$, ensuring that only the zero mode contributes significantly to the propagator for $pz,pz'\ll 1$.  It is the exponential suppression of the higher KK states near the Planck brane which guarantees that the one-loop RG flow of the Planck correlator is logarithmic.  For non-zero bulk mass $m$, a detailed analysis of the KK spectrum reveals the existence of an ``almost zero mode'' with 4D mass $m/\sqrt{2}$ whose wavefunction is unsuppressed near $z=1/k$ (the other KK modes are still peaked away from the Planck brane).  It is this mode which accounts for the pole in Eq.~(\ref{eq:asymp}).  See~\cite{pomarol,ads} for a discussion of this mode.

In terms of Eq.~(\ref{eq:asymp}), the $z$ integrals in Eq.~(\ref{eq:L1}) and Eq.~(\ref{eq:L2}) reduce to 
\begin{eqnarray}
\nn
  L^{(1)}_{\mu\nu} &=& \int {d^D p\over (2\pi)^D} {(2p+q)_\mu (2p+q)_\nu\over\left[(p^2+m_D^2) ((p+q)^2+m_D^2)\right]}\\
& & {}\times \left[{2k(\nu-1)\over q}\int_{1/k}^\infty dz (kz)^{D/2-2\nu} {K_{D/2-1}(qz)\over K_{D/2-2}(q/k)}\right]^2,
\end{eqnarray}
where we have defined $m_D^2=(1-2/D) m^2$ and neglected higher order terms in $q^2/k^2$.  Also,
\begin{equation}
L^{(2)}_{\mu\nu} = -2\eta_{\mu\nu}\int {d^D p\over (2\pi)^D} {1\over p^2 +m_D^2}\left[{2k(\nu-1)\over q}\int_{1/k}^\infty dz (kz)^{D-1-2\nu} \left({K_{D/2-1}(qz)\over K_{D/2-2}(q/k)}\right)^2\right],
\end{equation}
which again is only correct up to $q^2/k^2$ corrections.  Note that our approximation scheme, which uses the asymptotic propagator in the loop integrals,  has disentangled the original 5D loop diagram into the product of two simple integrals, one over the coordinates $z,z'$ with no dependence on the 4D loop momentum variable $p$, and a standard 4D one-loop integral.  This factorization holds up to power corrections in $q^2/k^2$.  Performing the $z$ integration we find, 
\begin{equation}
\label{eq:1l}
{g^2(q)\over q^2} = {g_5^2\over q} {K_{D/2-2}(q/k)\over K_{D/2-1}(q/k)}\left[1- {g_5^2} {q K_{D/2-2}(q/k)\over K_{D/2-1}(q/k)}\Pi(q^2)\right],
\end{equation}
where 
\begin{eqnarray}
\Pi(q^2)&=& {\Gamma(2-D/2)\over (4\pi)^{D/2}}\int_0^1 dx (1-2x)^2\left[x(1-x)q^2 + m_D^2\right]^{D/2-2},
\end{eqnarray} 
and we have dropped terms which are suppressed by powers of $k^{-1}$.   This result is ultraviolet divergent and must be subtracted via a suitable local counterterm to Eq.~(\ref{eq:1l}).  The functional dependence on $q/k$ suggests that the proper term to add arises from the contribution of a Planck brane localized kinetic operator for the gauge field\footnote{Note that we must add the local counterterm derived from a term in the $(D+1)$-dimensional action before taking the limit $D\rightarrow 4$.  Had we taken the 4D limit before subtracting the pole at $D=4$, the regularization procedure would have introduced spurious dependence on the external momentum that does not reflect the true non-analytic structure in the IR.  See~\cite{GnR0} for a more detailed discussion of dimensional regularization in warped background metrics.}
\begin{equation}
S_{bd} ={\lambda_k\over 4}\int d^D x F_{\mu\nu}(x,1/k) F^{\mu\nu}(x,1/k).
\end{equation}
It is easy to show that its contribution to Eq.~(\ref{eq:1l}) is
\begin{equation}
\left.{g^2(q)\over q^2}\right|_{\lambda_k} = - g_5^4\lambda_k\left({K_{D/2-2}(q/k)\over K_{D/2-1}(q/k)}\right)^2,
\end{equation}
which has the right dependence on $q/k$ to cancel the $1/(D-4)$ poles form $\Pi(q^2)$.  Thus we find, after resumming the one-loop result, and taking $q/k\ll 1$
\begin{equation}
\label{eq:pg}
{1\over g^2(q)} ={1\over g_5^2 k}\ln(k/q) + \lambda_k(\mu) -{1\over 48\pi^2}\log\left({q^2\over\mu^2}\right),
\end{equation}
where $\mu$ is a subtraction scale, and we have taken $m=0$ for simplicity (we have also dropped small, scheme-dependent constants). By choosing $\mu\simeq k$, such that the boundary coupling contains no large logarithms, we arrive at the usual 4D result for a massless scalar\footnote{Alternatively, one can take $\mu\simeq q$, in which case all the large logarithms are accounted for by the running boundary coupling $\lambda_k(q)$.}. The leading order tree level result is universal and irrelevant for the prediction of $\sin^2\theta_W$ in the context of a GUT. 

At this point one may wonder if this one-loop calculation is
meaningful.  After all, the tree-level propagator runs
logarithmically, so the one-loop running is,
strictly speaking, a pure counterterm.  One might conclude that the running
of a given gauge coupling is not calculable, since to determine the
beta function coefficient one must make a measurement of the 5D gauge
coupling $g_5$.  While it is true that the absolute running of the
Planck correlator is unobtainable from field theory, the relative running of two couplings is calculable if a symmetry relates the respective bare parameters. Of course, below the scale $T$, the running is again calculable since the classical logarithm  becomes independent of the momentum and serves only to define the value of the low energy tree-level coupling $g_4$.  Given an underlying GUT symmetry our procedure is in fact a way of isolating the parts of the 5D Feynman diagrams that do not cancel in coupling constant differences from those corrections that are purely universal and therefore unobservable.  Had we kept the UV modes with masses above $k$ in our calculation, they would simply generate a shift in the bulk coupling $g_5$, or in the value of the boundary coupling $\lambda_k(\mu)$.  However, these shifts would be unobservable if the symmetry breaking scale is below the curvature scale, which we implicitly assume here.

\subsection{Gauge Fields}
\label{sec:gauge}
\subsubsection{Running of gauge couplings}

The analysis of the one-loop contribution of bulk scalars to vacuum
polarization graphs was simplified due to the fact that, in terms of
KK modes, one can identify a single state which has non-negligible
overlap with the Planck brane, and thus gives rise to the leading
quantum effects.  The situation is more complicated in non-Abelian
gauge theories, where the KK excitations of the gauge fields appear in
the loops.  Then, it is no longer the case that a single mode
is responsible for the running of the couplings.  Rather, all KK
states with 4D masses up to a scale of order $k$ have non-negligible
wavefunction overlaps with the region $z=1/k$, and must be included in
the loops.  We will show that despite this fact, the leading
momentum dependence of gauge couplings as measured by Planck observers
is still logarithmic, with coefficients that identically match the 4D
beta function coefficients for non-Abelian gauge theories.

We begin by specifying the sense in which we use Planck correlators to
define effective gauge couplings in the non-Abelian case.  It is simplest
to define the couplings in terms of a gauge invariant Wilson loops that lie on the surface $z=1/k$.  These objects have exactly the properties we want: they are
Planck localized, and therefore calculable for a wide range of Wilson
loop sizes up to the length scale $1/k$.  They are also gauge
invariant, and can be used to extract physically meaningful
running couplings.  Thus we consider
\begin{equation}
\langle W(C)\rangle = \left\langle \mbox{Tr}\,{\cal P}\exp\left[ i \oint_C  dx^\mu A_\mu(x,1/k)\right]\right\rangle
\end{equation}
where $C$ is a closed loop that lies on the Planck brane, $z=1/k$.  It is convenient to compute this quantity by splitting up the bulk path integral measure as   
\begin{equation}
dA_M(X) = dA_\mu(x) d{\bar A}_M(X),
\end{equation}
where $A_\mu(x)=A_\mu(x,1/k)$ is the boundary value of the bulk gauge field, and ${\bar A}_M(X)$ is a bulk gauge field with boundary conditions on the Planck brane such that ${\bar A}_\mu(x,1/k)=A_\mu(x)$ (we denote full 5D coordinates by $X^M$, with $M$ running over $\mu,z$).  Then we may write the Wilson loop as 
\begin{equation}
\label{eq:PI4}
\langle W(C)\rangle =\int dA_\mu(x) \exp\left[-\Gamma[A(x)]\right]\mbox{Tr}\, {\cal P}\exp\left[ i \oint_C  dx^\mu A_\mu(x)\right],
\end{equation}
where 
\begin{equation}
\exp[-\Gamma[A(x)]] = \int d{\bar A}_M(X) \exp[-S[{\bar A}(X)]],
\end{equation}
with $S[A(X)]$ the classical gauge field action, including possible gauge fixing terms.  


By AdS/CFT, $\Gamma[A(x)]$ is the generating function for correlators of global symmetry currents $J^\mu_a(x)$ in the CFT dual to the AdS bulk theory
\begin{equation}
\exp[-\Gamma[A(x)]] = \left\langle\exp\left[i \int d^4 x A^a_\mu(x) J^\mu_a(x) \right]\right\rangle_{CFT}.
\end{equation}    
Inserting this into Eq.~(\ref{eq:PI4}), amounts to just a formal
derivation of the well known result that RS scenarios with bulk gauge
fields are dual to CFTs whose global symmetries are gauged.  We
conclude from Eq.~(\ref{eq:PI4}) that the value of $\langle
W(C)\rangle$, a 5D quantity, is obtained form a purely 4D computation
in which the propagators and vertices inserted into the graphs
generated by the functional integration over $A_\mu(x)$ are read from
the non-local 4D effective action $\Gamma[A(x)]$.

We construct $\Gamma[A(x)]$ including one-loop corrections in the AdS bulk as follows.  Let $A^{cl}_M(X)$ be an exact solution to the classical bulk Yang-Mills equations satisfying the specified boundary conditions.  We then expand the functional integration variable ${\bar A}$ as 
\begin{equation}
{\bar A}_M(X) = A^{cl}_M(X) + a_M(X),
\end{equation} 
where $a_M(X)$ is a quantum fluctuation about the background field which vanishes identically at $z=1/k$.  If we work to one-loop order, it suffices to keep terms up to quadratic order in $a_M(X)$.  In terms of this expansion, we have
\begin{equation}
\Gamma[A(x)] = S_{cl}[A(x)] + \delta\Gamma[A(x)],
\end{equation}  
where $S_{cl}[A(x)]$ is the bulk Yang-Mills action evaluated on the
classical solution $A_M^{cl}(X)$, and $\delta\Gamma[A(x)]$ represents
the result of integrating out the quantum fluctuation $a_M(X)$ at
one-loop.  Since $\delta\Gamma[A(x)]$ is already a one-loop quantity,
it only enters into the computation of $\langle W(C)\rangle$ as a
tree-level insertion into the functional integral over the boundary
field $A_\mu(x)$.  Therefore the quadratic piece in $A_\mu(x)$ of
$\delta\Gamma[A(x)]$ is sufficient for our purposes here.  However,
because $a_M(X)$ vanishes on the Planck brane, it is clear by the same type of arguments as in the previous section that the one-loop contribution of $a_M(X)$ does not give rise to logarithmic corrections: one can show that the classical field $A^{cl}_M(X)$ is
exponentially damped away from $z=1/k$  where the KK modes
of $a_M(X)$ have support.  Thus, $\delta\Gamma[A(x)]$ can at
most give rise to power corrections $q^2/k^2$, which we will simply drop from now on\footnote{In addition, note that if the GUT symmetry is broken only
on the boundaries, the vacuum polarization term $\delta\Gamma[A(x)]$
must be universal in the gauge indices, since the quadratic
fluctuation is in a complete multiplet of the bulk gauge group.  In
this case, the contribution of $\delta\Gamma[A(x)]$ to the Wilson loop cancels in the relative running between different gauge couplings.}, to the propagator for the boundary field $A_\mu(x)$.  The contribution from $\delta\Gamma[A(x)]$ to the Wilson loop is shown in Fig. 1a. The solid blob can be calculated via the diagrams in Fig. 1b and 1c.
\begin{figure}[ht]
\epsfxsize=10cm   
\centerline{\scalebox{0.6}{\includegraphics*[30, 340][595,555]{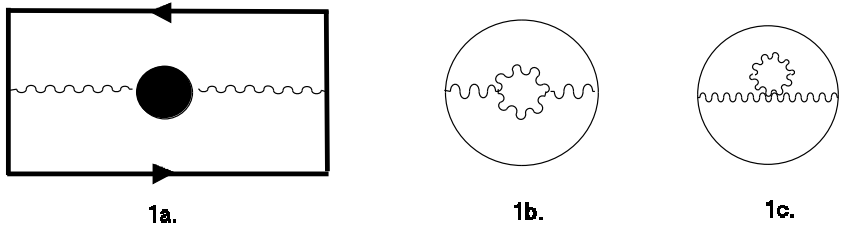}}}\caption{Fig. 1a, represents the 4D Wilson loop. The blob represents all the CFT corrections to the gauge boson two-point function. These corrections can be calculated using the AdS/CFT correspondence by calculating Witten diagrams. At the order we are working, we need to keep the Witten diagrams shown in Figs. 1b, 1c and 2b. The contribution from Fig. 2b gives the universal logarithm discussed in the text, while 1b and 1c represent the contributions from  $\delta\Gamma[A(x)]$ which are  possibly non-universal but power suppressed.}
\end{figure}
\begin{figure}[ht]
\epsfxsize=10cm   
\centerline{\scalebox{0.6}{\includegraphics*[30,340][595,555]{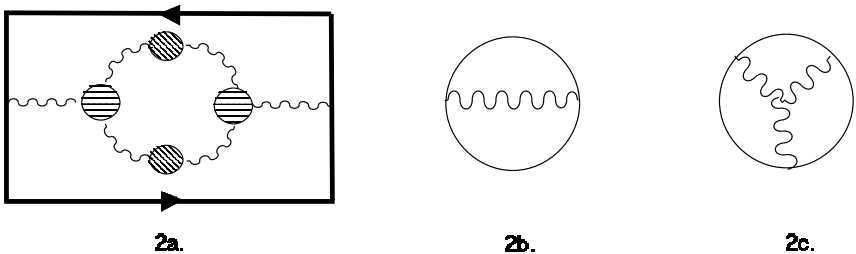}}}
\caption{Fig. 2a represents additional leading order corrections to the 4D Wilson loop. 2b, and 2c.are the Witten diagrams which would reproduce the two-point function represented by the filled circles in 2a.}
\end{figure}

It remains to compute the effective 4D propagator derived from $S_{cl}[A(x)]$ as well as the effective 4D vertices up to terms quartic in $A_\mu(x)$.  $S_{cl}[A(x)]$ also has terms higher order than quartic in $A_\mu(x)$ (these generate the higher $n$-point correlators of CFT currents), but we will not need these for our one-loop computation.  The necessary terms may be calculated by methods that are familiar from calculations of CFT current correlators via the AdS/CFT correspondence~\cite{AdSCFT,3pt}, and which are summarized in Fig. 2.  We will now show that despite the highly non-local nature of $S_{cl}[A(x)]$, the 4D one-loop computation of the Wilson loop in terms of the effective vertices (the blobs) in Fig. 2a can be reorganized into a term which leads to the usual running of the couplings in 4D gauge theory, and terms that are subleading in an effective expansion parameter $g^2(q)=g_5^2 k/\ln(k/q).$ 

First consider the quadratic terms in $A_\mu(x)$.  A standard computation gives
\begin{equation}
\label{eq:log}
S_{cl}[A(x)] ={1\over 2 g_5^2 k}\int {d^4 p\over (2\pi)^4} \log\left({\mu\over p}\right) A^a_\mu(p)\left[\eta^{\mu\nu} p^2 - p^\mu p^\nu\right] A^a_\nu(-p),
\end{equation}
resulting in the ``classical'' logarithmic running which we encountered previously.  Note that since we are interested in external momenta which are small compared to $k$ we have taken the position of the Planck brane to infinity (towards the AdS boundary).  The scale $\mu$ that appears in Eq.~(\ref{eq:log}) is an arbitrary subtraction scale which is necessary in order to define this singular limit.  Comparing with the form of the classical AdS gauge field propagator from Eq.~(\ref{eq:classrun}), we see that we must take $\mu$ of order the scale $k$ in order to reproduce the effects of cutting off AdS at $z=1/k$.  This is just the usual statement that truncating AdS with a UV brane at $z=1/k$ corresponds to cutting off the dual field theory at a scale $\mu\simeq k$ in the UV.  Note that this simple logarithmic dependence, identical in form to a one-loop 4D field theory calculation, follows from the fact that the classical running is generated by a two-point current correlator in a CFT, whose dependence on external momentum is uniquely fixed by the Ward identity and by conformal invariance.    

Unlike the two-point function, the three and four current correlators
are not uniquely determined by Ward identities and conformal
invariance.  This is easy to see at the level of the effective action
which generates the Green's functions of currents.  In a CFT, the two-point
function can only arise from a single non-local operator in the
effective action, that of Eq.~(\ref{eq:log}).  On the other hand, the
operators with more powers of $A_\mu(x)$ that generate $n$-point
functions may arise either from the terms related to
Eq.~(\ref{eq:log}) by gauge invariance, in the case of three and four
point functions, or by independent gauge-invariant operators with
three or more powers of the gauge field strength.

\begin{figure}[ht]
\epsfxsize=10cm   
\centerline{\scalebox{0.6}{\includegraphics*[20,350][600,625]{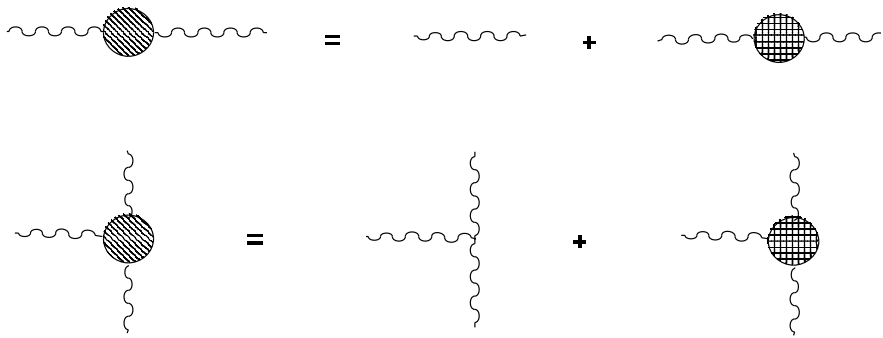}}}
\caption{The figure depicts the decomposition of the original non-local terms in the effective action (the blobs on the LHS) into local terms which define the fictitious coupling $g_F(\mu_0)$ and new non-local effective vertices (the blobs on the RHS).  An identical decomposition, not shown, is made for the four-point vertex.}
\end{figure}

Given the form of Eq.~(\ref{eq:log}), we are now in a position to see why the one-loop running of the gauge coupling defined through the Wilson loop is as in 4D.  It is useful to define a fictitious weak coupling by shifting the logarithmic term in the two-point function as\begin{equation}
{1\over g^2_5 k}\ln\left({k\over p}\right)\equiv {1\over g^2_F(\mu_0)} + {1\over g^2_5 k}\ln\left({\mu_0\over p}\right)
\end{equation} 
with $g^{-2}_F(\mu_0)= (g^2_5 k)^{-1}\ln(k/\mu_0)$.  Choosing $\mu_0$ to be near the scale of external momenta, the coupling $g_F(\mu_0)\ll 1$.  A similar decomposition can be made for the terms in the three and four point amplitudes whose non-analytic structure is related to the two-point function by the Ward identities.  See Fig. 3.  Thus, in terms of this coupling, we may rewrite the effective action as 
\begin{equation}
\label{eq:local}
S_{cl}[A(x)] = {1\over 4 g^2_F(\mu_0)}\int d^4 x F^a_{\mu\nu}(x) F_a^{\mu\nu}(x) + \int d^4 x A_\mu^a(x) J^\mu_a (x).
\end{equation}    
where the dynamical conserved current $J_\mu^a(x)$ mimics the relevant non-local effects of the $n$-point functions.  It is related to the CFT current by the addition of a local counterterm\footnote{Notice that the field redefinition necessary to get the normalization in Eq.~(\ref{eq:local}) is universal and does not affect physical predictions.}.  Besides the fact that the currents appear in complete multiplets of the GUT group and that they are conserved, the specific dynamics from which the $J^\mu_a(x)$ are constructed are not relevant to our discussion, and are ignored in Eq.~(\ref{eq:local}).  Since $g_F(\mu_0)$ is weak, it is now clear that to compute the one-loop logarithms, we may drop altogether insertions of $J^\mu_a(x)$, since at one-loop these are universal (reproducing the incalculable classical running in AdS) and beyond one-loop they are suppressed by powers of $\alpha_F(\mu_0)$ relative to the one-loop diagrams with the standard Yang-Mills propagators and vertices derived from the first term in Eq.~(\ref{eq:local}).  

More explicitly, consider for instance insertions of the three-point vertex into the 4D diagrams.  Provided $\mu_0$ is chosen to be of order the external momenta, it follows that the non-analytic terms in the vertex cannot contribute at leading log.  If $\mu_0$ is the scale of the external momenta, the diagrams of Fig. 2, with the blobs shifted as in Fig. 3, split up into a term which is just the usual 4D gauge theory graph, and a piece where the bare 4D graph is ``dressed'' by CFT propagator and vertex corrections.  The CFT corrections modify the loop integrals by logarithms of $l/\mu_0$, with $l$ some combination of loop and external momenta\footnote{There are also terms in the three-point function, arising from non-local operators with three powers of the gauge field strength, that do not have logarithms.  Such term, which are not constrained by the Ward identity, are not sensitive to the UV scales, and therefore cannot give additional large logs involving the scale $k$.}.  By dimensional analysis, the result of performing such integrals must yield logarithms of $q/\mu_0$, with $q$ the external momentum.  However, since $\mu_0\simeq q$, such logarithms are not large\footnote{Alternatively, we could have chosen, for example, $\mu_0\sim k$, in which case the logarithms are large.  Since $\mu_0$ is arbitrary, the results for a physical quantity cannot depend on how it is chosen.  Thus at leading log, the effects of shifting the value of $\mu_0$ can be absorbed into a universal redefinition of the tree level couplings.}.  We conclude from this that the non-analytic corrections to the local gauge theory vertices give rise to contributions to the running of the couplings which are down by a power of $g^2_F(\mu_0)$.  Identical statements can be made regarding insertions of the four-point function.  Thus, we see that at leading log, the non-universal contribution to the running of the Planck brane gauge couplings is identical to that found in a 4D gauge theory.  

\subsubsection{GUT predictions}

We are now in a position to apply these results to compute the running of the gauge couplings in a GUT model broken either by brane terms (either explicitly through orbifold boundary conditions or spontaneously by brane Higgs scalar fields) or by a bulk Higgs field.  First of all, it is clear that GUT breaking on the $z=1/T$ boundary cannot modify the running of the gauge couplings between the scales $k$ and $T$.  This is because for a wide range of external momenta $T\ll q\ll k$, TeV brane effects are exponentially damped by the external gauge boson propagators.  Thus we restrict our discussion to Planck brane or bulk breaking of the unified gauge group.  If the GUT symmetry is broken by adding a Planck localized 4D scalar field $\phi$ whose VEV breaks, say, $SU(5)$ down to $SU(3)\times SU(2)\times U(1)$, then it is obvious from Eq.~(\ref{eq:local}) that the dynamics is identical to that of a 4D gauge field coupled to the symmetry breaking $\phi$ field.  The contribution to the prediction for $\sin^2\theta_W$ from the gauge fields will involve the usual beta function coefficients that one would calculate in the Standard Model, multiplying large logarithms of $\langle\phi\rangle/m_W$, where $\langle\phi\rangle\leq k$ is the symmetry breaking VEV.  

Explicit breaking of $SU(5)$ by orbifold boundary conditions can be taken into account by inserting a delta functional into Eq.~(\ref{eq:PI4}) which enforces the vanishing of the components of the gauge field which are odd under the orbifold parity.  From the point of view of Eq.~(\ref{eq:local}) this case represents a situation in 4D where the full $SU(5)$ symmetry is never present at all.  Indeed, there are no dynamical gauge fields corresponding to the generators of $SU(5)$ which are not also in $SU(3)\times SU(2)\times U(1)$.  Boundary gauge kinetic terms at $z=1/k$ need not respect the full GUT structure of the bulk theory.  Since the addition of boundary gauge couplings corresponds to non-universal shifts in the UV values of the tree-level $\alpha_F^a(\mu_0)$, with $a=SU(3),SU(2),U(1)$, the couplings are never strictly unified\footnote{It is important to note, however, that the tree-level logarithmic running is still $SU(5)$ universal, so that the effects of explicit $SU(5)$ breaking appear only as non-universal boundary conditions for the RG running at a scale $\mu\simeq k$.}.   Thus, in order to make predictions, additional assumptions about the UV structure of the orbifold GUT are necessary.  It is customary to simply state~\cite{flat} that the boundary couplings at the scale $\mu\simeq k$ saturate the bound given by the assumption of strong coupling, in which case the uncertainty in the prediction for the low energy value of $\sin^2\theta_W$ due to the non-universality of the UV couplings is of order $\alpha_F(\mu_0)/(4\pi)$, and therefore suppressed relative to the leading log prediction.  Our discussion here is in quantitative agreement with the results of~\cite{choidec} obtained by summing KK one-loop corrections to the low-energy gauge couplings in orbifold GUTs.  However, our approach makes manifest the 4D nature of the prediction.

In the case where the GUT is broken spontaneously by the Higgs mechanism in the bulk space, the low energy couplings again obey the same relations that one would obtain by solving the RG equations for the non-Abelian gauge couplings in 4D.  In this case, the UV scale appearing in the logarithmic corrections to the gauge couplings is related to the bulk mass term acquired by the XY gauge bosons through the Higgsing of the gauge group, $m_{XY} \sim g_5\langle \Phi\rangle$ ($\Phi$ is a bulk scalar whose VEV breaks the GUT group to the Standard Model).  This can be seen most clearly by working in unitary gauge, where the relevant degrees of freedom are the massless bulk Standard Model gauge fields, the massive $XY$ bosons, and the massive Higgs field.  Assuming that all bulk masses are smaller than the curvature scale, it is clear from our analysis of scalar fields that the contribution of the Higgs to the running of the Planck brane gauge couplings is frozen out at a scale comparable to its bulk mass.  As is the case with massive scalar fields, the KK modes of a massive gauge field are localized away from the Planck brane, except for a single isolated mode with mass of order the bulk mass.  This can be seen by examining the pole structure of the massive gauge boson propagator.  Ignoring Lorentz structure, in $\mbox{AdS}_5$ this has the form 
\begin{equation}
D^{(v)}_p(z,z')\sim {(kz)^{1-\nu} (kz')^{1-\nu}\over p^2 + m^4/8k^2}, 
\end{equation}    
where we have taken $pz,pz'\ll 1$ (which as in the scalar case, is the relevant asymptotic region for obtaining the non-analytic effects).  Here, $\nu=\sqrt{1+m^2/k^2}$, with $m$ the gauge boson mass, which we have taken to be $m\ll k$.  From this equation, we see that the gauge bosons which acquire a mass due to the bulk Higgs drop out of the running at a scale $m_{GUT}\simeq m_{XY}^2/k\ll m_{XY}.$  If one is willing to tolerate a slight hierarchy in $m_{XY}/k$, then there may be interesting new model building possibilities.  For instance, one could have unification with an intermediate scale $m_{XY}^2/k$ without loss of predictivity.   This is in contrast to a typical two-scale scenario in 4D, where the intermediate scale introduces a new unknown parameter, thus weakening the theory's predictive power.  

If GUT breaking occurs at a scale such that $m_{XY}>k$, then the dominant corrections to the low energy gauge couplings can be obtained by a flat spacetime computation.  For $q\gg k$, the Planck correlation function that defines the running gauge coupling exhibits power law running, just like in flat 5D space.  This running is universal, until $q\sim m_{XY}$, at which point the heavy gauge bosons drop out of the running and only the Standard Model gauge fields contribute.  The effect of integrating out the heavy fields can be incorporated into a non-universal shift in the bulk gauge coupling
\begin{equation}
{1\over g_5^2}\rightarrow {1\over g_5^2} + m_{XY},
\end{equation}
with a different group theory coefficient multiplying $m_{XY}$ for each factor of the Standard Model group not shown (we have also dropped factors of $1/16\pi^2$ that arise from the loop integration) .  When evaluated at a scale $q\ll k$, the Planck gauge coupling will thus receive, in light of Eq.~(\ref{eq:classrun}), a non-universal logarithmic correction  
\begin{equation}
{1\over g^2(q)} \sim {m_{XY}\over k}\log(k/q) + \mbox{universal}. 
\end{equation}
where we have dropped the logarithmic one-loop corrections of the
massless gauge fields relative to the power enhanced term.  When
evaluated at $q\sim T$, this gives rise to a correction to the low
energy gauge couplings of the form $m_{XY} R$.  Thus Planck
correlators reproduce power law effects when the symmetry breaking
scale lies above $k$.  The limit $m\gg k$ has also been discussed
in~\cite{ads,GnR2}.  Such power law terms invalidate the systematics,
which is why we restrict ourselves to Higgs breaking scales below the
curvature scale, so that the leading quantum corrections are only
logarithmically sensitive to the UV scale.

\subsection{Fermions}
\label{sec:fermions}

To compute Planck brane vacuum polarization corrections due to bulk fermions, we proceed as in the scalar case considered in Sec.~\ref{sec:scalar}.  We will need the asymptotic form of the bulk propagators in the limit $pz,pz'\ll 1$ appropriate for our choice of scheme.  Using the results of the appendix, we find in this limit, for $m/k>1/2$,
\begin{equation}
\label{eq:mlarge}
D^{(f)}_p(z,z') = -i P_+ {\pslash\over p^2} \left(2m -k\right) (kz)^{D/2-m/k-1} (kz')^{D/2-m/k-1}, 
\end{equation}
where we have only shown the term whose Dirac structure leads to a non-vanishing contribution to the vacuum polarization.  Note that we have taken fermion boundary conditions such that the right handed components of the bulk fermion are non-vanishing on the Planck brane.  In this equation, $P_+=(1+\gamma_5)/2$.  For $m/k=1/2$, 
\begin{equation}
D^{(f)}_p(z,z') = - i P_+ {\pslash}{k\over p^2 \ln\left({2k\over p}\right)} (kz)^{D/2-3/2} (kz')^{D/2-3/2},
\end{equation}
and for $m/k<1/2$, we have
\begin{equation}
\label{eq:msmall}
D^{(f)}_p(z,z')=-i P_+{\pslash\over k}{\Gamma(m/k+1/2)\over\Gamma(1/2-m/k)} \left({2k\over p}\right)^{1+2m/k}(kz)^{D/2-m/k-1} (kz')^{D/2-m/k-1}.
\end{equation}
These results are easy to understand in terms of the KK structure of bulk fermions in the truncated $\mbox{AdS}_5$ background~\cite{grossman}.  The KK spectrum of a bulk fermion contains a 4D chiral zero mode whose degree of overlap with the Planck brane depends on the ratio $m/k$.  In the case $m/k<1/2$ this chiral zero mode, as well as the KK modes, are localized to the TeV brane.  For $m/k>1/2$, the chiral mode is localized towards the Planck brane, while the KK modes remain on the TeV brane.  This accounts for the isolated massless pole in Eq.~(\ref{eq:mlarge}).  In the case with $m=k/2$, the chiral zero mode is flat and has order unity overlap with the Planck brane, but the KK modes are not exponentially suppressed there, leading to the logarithmic enhancement of the propagator.  In fact, supersymmetry in AdS fixes the mass of the bulk fermion which appears in a gauge multiplet to be $m=k/2$, so the $m=k/2$ fermion KK modes are identical to the wavefunctions of a massless bulk gauge field~\cite{gherghetta}.

From these propagators, we naively conclude that the cases in which $m/k\geq 1/2$ generate the same large non-universal logarithm in the predictions for low energy gauge couplings as does a single chiral fermion in 4D.  For $m/k>1/2$ this is a consequence of the dominance of the chiral mode contribution to the Planck correlator, while for $m=k/2$ one can use the same type of arguments as in the previous section to show that the logarithmic corrections to the propagator do not modify the leading log prediction that one gets by just keeping the zero mode.  Despite the presence of a massless chiral mode in the low energy spectrum, no such logarithm arises for $m<k/2$, since the propagator Eq.~(\ref{eq:msmall}) is more singular in the UV.  

Note however, that some care must be taken in interpreting the fermionic corrections to low energy gauge couplings.  First of all, the statements of the previous paragraph would seem to indicate that the fermionic corrections are discontinuous functions of the bulk mass parameter $m$ as it crosses $m/k=1/2$ .  This is because our scheme actually misses power corrections of the form $(m/k) \ln(q/k)$ which are negligible for $m/k\ll 1$, but become as large as the zero mode logarithms for $m/k$ of order unity.  Including these terms restores the continuity of the one-loop result as a function of the bulk mass.  However, as already discussed in the previous section, when $m$ becomes larger than order $k$ these power corrections arise from purely flat spacetime effects, and correspond to non-universal shifts in the values of the bulk gauge couplings.  These effects are therefore not accessible to the Planck brane scheme, which treats the UV modes above the scale $k$ as universal.  In any case, at low energy these power corrections give rise to a contribution proportional to $m R$, with $R=k^{-1} \ln(k/T)$, which is just the usual power law running effect encountered in 5D flat space.  Such linear sensitivity to the UV jeopardizes the possibility of making robust predictions for physical observables such as $\sin^2\theta_W$ in GUT models.  The explicit linear dependence on the fermion bulk mass parameter is borne out by the results of~\cite{choidec}.

Since there is no chiral symmetry to protect the fermion bulk mass
 from becoming of order the Planck scale in five dimensions, it would
 seem that bulk fermions are irrelevant. When $m<k$ there is no effect
 on the running, and when $m>k$ the systematics are seemingly lost.
 There are situations, however, where the fermionic power corrections
 are under control.  For instance, if the GUT is broken in such a way
 that the bulk mass parameters remain universal (as typically happens
 in models where the symmetry is broken by boundary effects only) then
 the incalculable power corrections cancel in the prediction for the
 weak mixing angle.  In addition, if the GUT is broken by orbifold
 boundary conditions, the zero mode structure need not appear in complete multiplets, in which case the large logarithms that arises if $m/k\geq 1/2$ are meaningful, since the UV sensitive power piece cancels in the prediction\footnote{For instance, in the supersymmetric GUT model of~\cite{gns}, the MSSM Higgs doublets arise as the zero modes that remain from the orbifold projection of bulk $\bf 5$ and $\bf\bar 5$ Higgs hypermultiplets.  As long as the bulk masses of these hypermultiplets are larger than or equal to $k/2$, the doublet zero modes contribute large logarithms to $\sin^2\theta_W$ just as in the MSSM.}.  Furthermore, in supersymmetric models, the mass of a bulk hypermultiplet does not receive radiative corrections, so small  fermion masses are technically natural.  Thus in this case, although  the chiral zero mode does not contribute to the running of the Planck  gauge couplings, the power corrections can be made small and under control.

We note also that from the perspective of AdS/CFT, a $m/k\gg 1$ fermion field whose zero mode is not projected out by boundary conditions will have, in the CFT description, a corresponding fundamental field which couples to an irrelevant CFT operator of conformal dimension $2+m/k$~\cite{spinor}. Thus, the prediction for the running of the gauge coupling in the 4D dual CFT will be just the 4D logarithm of a chiral fermion.  The large power law corrections cannot be seen in the CFT since they correspond to effects that are purely Minkowskian.

\section{Tools for GUT model building}

This self-contained section is intended to be a summary of our results for model builders who would like to calculate the prediction for $\sin^2\theta_W$ in a given model.  A discussion of non-supersymmetric warped GUTs appears in~\cite{ads1}.  GUT model building in the context of warped supersymmetry has been discussed in~\cite{pomarol,gns,Chacko,Hall}.

In building models one may choose to break the GUT symmetry spontaneously, via Higgs fields in the bulk or on the branes, or explicitly through orbifold boundary conditions.  One may also add boundary kinetic and mass terms for the bulk fields, as well as fields that are confined to propagate strictly on the boundaries.  In the Higgs case, the boundary terms must be consistent with the underlying gauge symmetry, and the mass term arises from the spontaneous breaking.  On the other hand, in the orbifold case, the brane actions need not respect the GUT symmetry if the boundary conditions do not.  For the purposes of computing the leading log corrections to an observable like $\sin^2\theta_W$, one may ignore completely the details of the TeV brane dynamics.  This is clear from the point of view of Planck correlators, as discussed in the previous sections and also in~\cite{GnR1,GnR2}.

The analysis of Planck correlators makes it clear that, whenever calculable, the one-loop quantum corrections to the low energy gauge couplings contain large logarithms whose coefficients can be read off the standard 4D beta function coefficients for the bulk field KK zero modes which have support on the Planck brane.  There will also be scheme dependent, universal, constants which will cancel in predictions for coupling constant differences.  For instance, in the model of~\cite{gns}, the 4D zero modes consist of the MSSM vector and chiral multiplets, with the $SU(2)$ Higgs doublets arising as Planck brane localized ($m/k\geq 1/2$) zero modes of bulk hypermultiplets.  The predictions for $\sin^2\theta_W$ in that model are identical to those in the ordinary (4D) MSSM, despite the 5D nature of the model for energies higher than the KK mass gap.  It is important to note that this simple rule for computing one-loop corrections assumes that non-universal bulk masses, when they arise, are smaller than the curvature scale $k$.  Symmetry breaking masses larger than the curvature scale lead to large power corrections to the low energy gauge couplings whose origin is identical to the power law running encountered in flat space higher dimensional gauge theories.  Because of the high sensitivity to UV effects, such power corrections lead to a loss of predictivity in the low energy observables.  Thus we will only consider situations in which the GUT breaking occurs at a scale smaller than the curvature scale.  Finally, it is clear that 4D fields confined to the Planck brane contribute the usual 4D logs to the running of the couplings, while TeV brane fields do not contribute large logarithms at all.

While the running of the couplings is effectively determined by the zero mode structure of a given model, the UV scale (i.e. the GUT scale) appearing in the low energy predictions is dependent on the manner in which the GUT symmetry is broken.  For orbifold breaking on the Planck brane, there is, strictly speaking, no GUT scale.  Indeed, there are tree-level contributions to the gauge couplings due to Planck brane localized gauge kinetic operators for the Standard Model gauge fields which do not respect the full GUT symmetry. To retain predictivity at one-loop, one assumes~\cite{flat} that the non-universal boundary couplings evaluated at a scale near the 5D Planck scale, where the GUT becomes strongly coupled, are of order $1/16\pi^2$, and therefore suppressed relative to the logarithmically enhanced one-loop corrections.  For a discussion of this in the context of AdS, see~\cite{GnR2,choidec}.  Thus, under this strong coupling assumption, the argument of the logarithms in the one-loop prediction for $\sin^2\theta_W$ in a warped GUT broken by Planck brane orbifold boundary conditions is of order $m_W/M_5$.  The coefficient of the logarithm is determined solely by the bulk field zero modes with support on the Planck brane.  There are uncertainties in this prediction which arise due to the unknown values of the boundary gauge couplings, and which have magnitude comparable to two-loop effects in 4D gauge theory.

If the GUT is broken spontaneously on the Planck brane, then the UV scale appearing in the logarithmic correction to the low energy couplings is the symmetry breaking VEV $\langle\phi\rangle\leq k$.  Thus the low energy $\sin^2\theta_W$ prediction will again contain logarithms involving a large scale.  In non-supersymmetric models it is also possible to break the GUT symmetry spontaneously in the bulk by the Higgs mechanism.  While in this case the running of the couplings is largely determined by the zero modes, the low energy predictions may differ from what one finds in the Standard Model.  This is because, as discussed in the previous section, the XY gauge bosons drop out of the running of the gauge couplings at a scale $g_5^2\langle\Phi\rangle^2/k$, which is lower than the scale $g_5\langle\Phi\rangle$ where the Higgs fields responsible for breaking the GUT to the Standard Model gauge group stop their contribution to the running (here $\langle\Phi\rangle$ is the bulk Higgs VEV).  Given that we take $g_5\langle\Phi\rangle\ll k$ to avoid large non-universal power corrections, it follows that there may be significant, but model dependent, corrections to $\sin^2\theta_W$ from the running between $g_5\langle\Phi\rangle$ and $g_5^2\langle\Phi\rangle^2/k$.  This deviation from the 4D Standard Model result is in addition to the small warped space corrections of the form $g_5^2\langle\Phi\rangle^2\ln(k/T)/16\pi^2 k^2$ emphasized in~\cite{ads1}.  Also, in models where the warped geometry addresses the hierarchy problem, the Standard Model Higgs doublet is confined to TeV brane, and therefore does not contribute large logarithms to the running of the gauge couplings~\cite{GnR1,GnR2}.  It is therefore possible that these combined effects improve the UV convergence of the gauge couplings relative to the situation in the minimal Standard Model.  Unfortunately, in models with gauge bosons in the bulk and the Standard Model Higgs on the TeV brane there are large tree-level corrections to precision electroweak observables~\cite{cet} unless the TeV brane scale $T$ is an order magnitude larger than the TeV scale.  Thus, further model building is still needed to make non-supersymmetric warped GUTs viable.

\section{Summary and Conclusions}

A common well founded criticism of RS type scenarios is based on the compelling unification of the gauge couplings in the MSSM. At first sight it would seem that any new physics due to hidden dimensions at scales below the GUT scale would ruin unification. For flat manifolds this is indeed true. However, if the compactified bulk is AdS, logarithmic unification is not lost.  Thus it is possible to keep the usual 4D GUT prediction for $\sin^2\theta_W$ and yet still observe extra dimensions at the TeV scale.  In particular, it is possible that one could observe the KK modes of $X,Y$ bosons at this scale and still not violate proton decay bounds~\cite{pomarol,gns}.

In this paper we have given a complete account of the systematics for the running of gauge couplings in compactified $\mbox{AdS}_5$, including the one-loop effects of bulk scalar, fermion, and vector fields.  In particular, we have explained how this miracle of effective 4D running can occur in terms of effective field theory arguments.  From our analysis of Planck brane gauge correlators emerges a simple set of rules for determining how the 5D field content of a given model contributes to the running of the couplings.  In a large class of models, one finds that bulk fields with zero modes with support on the Planck brane will give rise to logarithms as in 4D field theories.  We have also discussed the UV scales that appear in the predictions for $\sin^2\theta_W$ when the warped GUT is broken by either bulk Higgs fields or by boundary (orbifold or Higgs) effects.  In the bulk Higgs case, we find that the effects of massive bulk gauge bosons decouple at a scale which is parametrically suppressed by a power of $m/k$ relative to their naive bulk mass $m$.  This could allow for the possibility of building models with two-step unification without loss of predicitivity, if one is willing to tolerate a slight hierarchy in $M_{GUT}/k$.  Finally, we note that the mechanism employed here for understanding how the gauge couplings run in AdS should also be applicable to determining the RG flows of other bulk couplings (for instance Yukawa couplings) relevant to building models of grand unification based on the AdS warp factor.

\section{Acknowledgments}
W.G. is supported in part by the DOE contract DE-AC03-76SF00098 and by the NSF grant PHY-0098840.  The work of I.R. is supported in part by the DOE contracts DOE-ER-40682-143 and DEAC02-6CH03000.

\appendix

\section{Propagators}
\label{app:props}
Here we write down the expressions for the bulk field propagators used in the text.  

\subsection{Scalars}
In the mixed $(p_4,z)$ representation, the scalar propagator satisfies the equation
\begin{equation}
\left[(kz)^{(D-2)} \partial_z\left({1\over (kz)^{(D-2)}}\partial_z\right) -p^2 -{m^2\over(kz)^{(D-3)}}\right]D_p(z,z') = -(kz)^{(D-2)}\delta(z-z')
\end{equation}
As should be clear from the discussion in the text, the presence of the TeV brane is irrelevant as far as understanding the large logarithms that arise in low energy gauge coupling predictions.  It will therefore suffice to calculate propagators in the limit $T\rightarrow 0$.  Thus for the scalar propagator we impose
\begin{equation}
D_p(z\rightarrow\infty,z')=0.
\end{equation}
On the Planck brane we will consider either Neumann ($\partial_z\Phi=0$) or Dirichlet ($\Phi$=0) scalar field boundary conditions.  For Neumann conditions on the Planck brane, the scalar propagator is
\begin{equation}
D_p(z,z')= -{1\over k}{(kz)^{D/2} (kz')^{D/2}\over k_\nu(p/k)}\left[i_\nu(p/k) K_\nu(pz)-k_\nu(p/k) I_\nu(pz)\right]K_\nu(pz')
\end{equation}
while for $z\geq z'$ we simply switch the two factors.  In this equation, we have defined 
\begin{equation}
i_\nu(z) = z^{1-D/2} {d\over dz}\left(z^{D/2} I_\nu(z)\right),
\end{equation}
and a similar expression for $k_\nu(z)$ with $I_\nu\rightarrow K_\nu$ ($I_\nu(z), K_\nu(z)$ are the modified Bessel functions).  The scalar mass parameter enters this expression through $\nu=\sqrt{(D/2)^2 +m^2/k^2}$.  The Dirichlet propagator is given by 
\begin{equation}
D_p(z,z')= -{1\over k}{(kz)^{D/2} (kz')^{D/2}\over K_\nu(p/k)}\left[I_\nu(p/k) K_\nu(pz)-I_\nu(p/k) I_\nu(pz)\right]K_\nu(pz')
\end{equation}

\subsection{Fermions}

There is a choice of spin connection for which the propagator of a Euclidean fermion with action 
\begin{equation}
S=\int d^{D+1} X\sqrt{G} {\bar \psi} \left( \Dslash -m \right)\psi 
\end{equation}
is given by, in mixed position/momentum
\begin{equation}
\label{eq:fprop}
\left[i\pslash -\gamma_5\left(\partial_z -{D\over 2z}\right) -{m\over (kz)}\right]D^{(f)}_p(z,z') = (kz)^{D-1}\delta(z-z')
\end{equation}
Let $P_{\pm} =(1\pm\gamma_5)/2$.  Then the solution of this equation can be written as
\begin{equation}
D^{(f)}_p(z,z') = \sum_{i\in\{+,-\}} P_i\left[S^i (z,z') + \pslash V^i(z,z')\right]. \end{equation}
Where Eq.~(\ref{eq:fprop}) imposes the relations
\begin{equation}
S^{\pm} = \pm i \left(\partial_z - {D\over 2 z}\mp m\right) V^\mp.
\end{equation}
Without loss of generality, we can impose boundary conditions on the bulk spinor such that the $P_-\psi(z=1/k,x)=0$.  In that case $S^-(z=1/k,z')=V^-(z=1/k,z')=0$, and the functions $V^{\pm}(z,z')$ are given by
\begin{equation}
V^+(z,z') = -{i\over k} {(kz)^{(D-1)/2}(kz')^{(D-1)/2}\over K_{\nu_+-1}(p/k)}\left[K_{\nu_+-1}(p/k) I_{\nu_+}(pz) + I_{\nu_+-1}(p/k) K_{\nu_+}(pz)\right] K_{\nu_+}(pz'),
\end{equation}
and
\begin{equation}
V^-(z,z') = {i\over k} {(kz)^{(D-1)/2}(kz')^{(D-1)/2}\over K_{\nu_-}(p/k)}\left[K_{\nu_-}(p/k) I_{\nu_-}(pz) - I_{\nu_-}(p/k) K_{\nu_-}(pz)\right] K_{\nu_-}(pz'),
\end{equation}
where $\nu_\pm=\left|m/k\pm1/2\right|$.   Note that for $m=0$ these expressions coincide with the propagator for a massless fermion in flat space.  This is easily seen by making the field redefinition $\psi(x,z)\rightarrow(kz)^{D/2}\psi(x,z)$ and ${\bar \psi}(x,z)\rightarrow (kz)^{D/2}{\bar\psi(x,z)}$, which removes both the spin connection term and the factor of $\sqrt{G}$ in the action.

\subsection{Vectors}

The quadratic part of the gauge field action can be written as 
\begin{eqnarray}
\nonumber
S_2 &=& {1\over 2}\int d^{D} x {dz\over (kz)^{D-3}}\left[-A_\mu\left(\eta^{\mu\nu}\partial^2 + \eta^{\mu\nu}(kz)^{D-3}\partial_z\left({1\over (kz)^{D-3}}\partial_z\right)-\partial^\mu\partial^\nu\right)A_\nu\right. \\
& & \left. {} + 2\eta^{\mu\nu} A_z\partial_z\partial_\mu A_\nu -A_z \partial^2 A_z\right]
\end{eqnarray}
It is convenient to remove the mixing between $A_z, A_\mu$ by adding a gauge fixing term of the form 
\begin{equation}
S_{GF} = {1\over 2\xi} \int d^D x {dz\over (kz)^{D-3}}\left[\partial_\mu A^\mu - \xi (kz)^{D-3}\partial_z\left({1\over(kz)^{D-3}} A_z\right)\right]^2. 
\end{equation}
Then the classical $A_\mu$ propagator satisfies
\begin{equation}
\left[-\eta^{\mu\alpha}p^2 +\eta^{\mu\alpha}(kz)^{D-3} \partial_z\left({1\over (kz)^{D-3}}\partial_z\right) + \left(1-{1\over\xi}\right)p^\mu p^\alpha\right]D_{p;\alpha\nu}(z,z') = -(kz)^{D+1}\delta^\mu_\nu\delta(z-z'). 
\end{equation}
We will work in Landau gauge, $\xi=\infty$.  Then the gauge boson propagator is
\begin{equation}
D_{p;\mu\nu}(z,z') = \left(\eta_{\mu\nu} - {p_\mu p_\nu\over p^2}\right) D^{(v)}_p(z,z').
\end{equation}  
For Neumann boundary conditions on the Planck brane this is ($z<z'$)
\begin{eqnarray}
\nn
D^{(v)}_p(z,z') &=& {1\over k}{(kz)^{D/2-1} (kz')^{D/2-1}\over K_{D/2-2}(p/k)} \left[K_{D/2-2}(p/k) I_{D/2-1}(pz)\right. \\
               & & \left.{} + I_{D/2-2}(p/k) K_{D/2-1}(pz)\right] K_{D/2-1}(pz')
\end{eqnarray}



\begin{thebibliography}{}
\bibitem{RS1}
L.~Randall and R.~Sundrum,
Phys.\ Rev.\ Lett.\  {\bf 83}, 3370 (1999)
[arXiv:hep-ph/9905221].

\bibitem{gw}
W.~D.~Goldberger and M.~B.~Wise,
Phys.\ Rev.\ Lett.\  {\bf 83}, 4922 (1999)
[arXiv:hep-ph/9907447].

\bibitem{pomarol}
A.~Pomarol,
Phys.\ Rev.\ Lett.\  {\bf 85}, 4004 (2000)
[arXiv:hep-ph/0005293].

\bibitem{GnR1}
W.~D.~Goldberger and I.~Z.~Rothstein,
Phys.\ Rev.\ Lett.\  {\bf 89}, 131601 (2002)
[arXiv:hep-th/0204160].

\bibitem{maldacena}
J.~Maldacena,
Adv.\ Theor.\ Math.\ Phys.\  {\bf 2}, 231 (1998)
[Int.\ J.\ Theor.\ Phys.\  {\bf 38}, 1113 (1998)]
[arXiv:hep-th/9711200].

\bibitem{AdSCFT}
S.~S.~Gubser, I.~R.~Klebanov and A.~M.~Polyakov,
Phys.\ Lett.\ B {\bf 428}, 105 (1998)
[arXiv:hep-th/9802109].
E.~Witten,
Adv.\ Theor.\ Math.\ Phys.\  {\bf 2}, 253 (1998)
[arXiv:hep-th/9802150].

\bibitem{Gubser}
S.~S.~Gubser,
Phys.\ Rev.\ D {\bf 63}, 084017 (2001)
[arXiv:hep-th/9912001].

\bibitem{APR}
N.~Arkani-Hamed, M.~Porrati and L.~Randall,
JHEP {\bf 0108}, 017 (2001)
[arXiv:hep-th/0012148].

\bibitem{RZ}
R.~Rattazzi and A.~Zaffaroni,
JHEP {\bf 0104}, 021 (2001)
[arXiv:hep-th/0012248].

\bibitem{PV}
M.~Perez-Victoria,
JHEP {\bf 0105}, 064 (2001)
[arXiv:hep-th/0105048].

\bibitem{ads}
K.~Agashe, A.~Delgado and R.~Sundrum,
Nucl.\ Phys.\ B {\bf 643}, 172 (2002)
[arXiv:hep-ph/0206099].

\bibitem{cct}
R.~Contino, P.~Creminelli and E.~Trincherini,
JHEP {\bf 0210}, 029 (2002)
[arXiv:hep-th/0208002].

\bibitem{GnR2}
W.~D.~Goldberger and I.~Z.~Rothstein,
arXiv:hep-th/0208060.

\bibitem{choidec}
K.~w.~Choi and I.~W.~Kim,
arXiv:hep-th/0208071.

\bibitem{ads1}
K.~Agashe, A.~Delgado and R.~Sundrum,
arXiv:hep-ph/0212028.

\bibitem{rschw}
L.~Randall and M.~D.~Schwartz,
Phys.\ Rev.\ Lett.\  {\bf 88}, 081801 (2002)
[arXiv:hep-th/0108115].
L.~Randall and M.~D.~Schwartz,
JHEP {\bf 0111}, 003 (2001)
[arXiv:hep-th/0108114].

\bibitem{choi}
K.~w.~Choi, H.~D.~Kim and Y.~W.~Kim,
arXiv:hep-ph/0207013.

\bibitem{decon}
A.~Falkowski and H.~D.~Kim,
JHEP {\bf 0208}, 052 (2002)
[arXiv:hep-ph/0208058];
L.~Randall, Y.~Shadmi and N.~Weiner,
JHEP {\bf 0301}, 055 (2003)
[arXiv:hep-th/0208120].


\bibitem{ad}
K.~Agashe and A.~Delgado,
arXiv:hep-th/0209212.

\bibitem{GnR0}
W.~D.~Goldberger and I.~Z.~Rothstein,
Phys.\ Lett.\ B {\bf 491}, 339 (2000)
[arXiv:hep-th/0007065].

\bibitem{3pt}
D.~Z.~Freedman, S.~D.~Mathur, A.~Matusis and L.~Rastelli,
Nucl.\ Phys.\ B {\bf 546}, 96 (1999)
[arXiv:hep-th/9804058];
G.~Chalmers, H.~Nastase, K.~Schalm and R.~Siebelink,
Nucl.\ Phys.\ B {\bf 540}, 247 (1999)
[arXiv:hep-th/9805105].

\bibitem{flat}
L.~J.~Hall and Y.~Nomura,
Phys.\ Rev.\ D {\bf 64}, 055003 (2001)
[arXiv:hep-ph/0103125];
Phys.\ Rev.\ D {\bf 65}, 125012 (2002)
[arXiv:hep-ph/0111068];
Y.~Nomura, D.~R.~Smith and N.~Weiner,
Nucl.\ Phys.\ B {\bf 613}, 147 (2001)
[arXiv:hep-ph/0104041];
R.~Contino, L.~Pilo, R.~Rattazzi and E.~Trincherini,
Nucl.\ Phys.\ B {\bf 622} (2002) 227
[arXiv:hep-ph/0108102].

\bibitem{grossman}
Y.~Grossman and M.~Neubert,
Phys.\ Lett.\ B {\bf 474}, 361 (2000)
[arXiv:hep-ph/9912408].

\bibitem{gherghetta}
T.~Gherghetta and A.~Pomarol,
Nucl.\ Phys.\ B {\bf 586}, 141 (2000)
[arXiv:hep-ph/0003129].

\bibitem{gns}
W.~D.~Goldberger, Y.~Nomura and D.~R.~Smith,
arXiv:hep-ph/0209158.

\bibitem{spinor}
M.~Henningson and K.~Sfetsos,
Phys.\ Lett.\ B {\bf 431}, 63 (1998)
[arXiv:hep-th/9803251];
W.~Muck and K.~S.~Viswanathan,
Phys.\ Rev.\ D {\bf 58}, 106006 (1998)
[arXiv:hep-th/9805145].

\bibitem{Chacko}
Z.~Chacko and E.~Ponton,
arXiv:hep-ph/0301171.

\bibitem{Hall}
L.~J.~Hall, Y.~Nomura, T.~Okui and S.~J.~Oliver,
arXiv:hep-th/0302192.

\bibitem{cet}
C.~Csaki, J.~Erlich and J.~Terning,
Phys.\ Rev.\ D {\bf 66}, 064021 (2002)
[arXiv:hep-ph/0203034].



\end{thebibliography}
\end{document}